\documentclass[aps,twocolumn]{revtex4}
\usepackage{graphicx}
\usepackage{subfigure}

\usepackage{amsmath}

\begin{document}

%%%%%%%%%%%%%%%%%%%%%%%%%%%%%%%%%%%%%%%%%%%%%%%%%%%%%%%%%%%%%%%%%%%%%%%%%%%
%%%%%%%%%%%%%%%%%%%%%%%%%%%%%%%%%%%%%%%%%%%%%%%%%%%%%%%%%%%%%%%%%%%%%%%%%%%
 
\title{Transport and Strong-Correlation Phenomena in \\
 Carbon Nanotube Quantum Dots in a Magnetic Field}
 \author{M. Mizuno$^1$, Eugene H. Kim$^1$ and G.~B. Martins$^2$}
\affiliation{$^1$ Department of Physics, University of Windsor,
         Windsor, Ontario, Canada N9B 3P4  \\
	 $^2$ Department of Physics, Oakland University, 
	 Rochester, Michigan 48309}

\begin{abstract}
Transport through carbon nanotube (CNT) quantum dots (QDs) in 
a magnetic field is discussed.
The evolution of the system from the ultraviolet to the 
infrared is analyzed; 
the strongly correlated (SC) states arising in the infrared 
are investigated.
Experimental consequences of the physics are presented ---
the SC states arising at various fillings are shown to be 
drastically different, with distinct signatures in the 
conductance and, in particular, the noise.
Besides CNT QDs, our results are also relevant to double QD systems. 
\end{abstract}

% \pacs{71.27.+a,72.15.Qm,73.63.-b,73.63.Kv}
\maketitle

%%%%%%%%%%%%%%%%%%%%%%%%%%%%%%%%%%%%%%%%%%%%%%%%%%%%%%%%%%%%%%%%%%%%%%%%%%%
%%%%%%%%%%%%%%%%%%%%%%%%%%%%%%%%%%%%%%%%%%%%%%%%%%%%%%%%%%%%%%%%%%%%%%%%%%%

% \section{Introduction}

Since their discovery, carbon nanotubes (CNTs) have been the subject of 
intense activity;\cite{tubereview}
in particular, experiments on transport in CNTs have revealed
a wealth of exciting phenomena.
Indeed, long metallic CNTs have been shown to behave as quantum 
wires;\cite{wire,postma} 
negative differential resistance has been observed in semiconducting 
CNTs.\cite{sc} 
Furthermore, short CNTs have been shown to behave as quantum dots 
(QDs),\cite{tubedot,postma,tubekondo} exhibiting Coulomb blockade (CB) 
phenomenology\cite{coulomb} known from gated two-dimensional semiconducting 
structures.

QDs have spurred a renewed excitement about the Kondo effect (KE), 
as they allow detailed 
investigations of the phenomena.\cite{leo}  
In this regard, CNT QDs are ideal for studies of Kondo physics. 
Indeed, initial experiments displayed an SU(2) KE arising from
the electron's spin;\cite{tubekondo}
more recently, orbital\cite{jarillo} as well as SU(4) KEs 
have been observed.\cite{jarillo,duke1,duke2}
Furthermore, CNT QDs afford the possibility of tuning between 
a variety of strongly-correlated (SC) states with a magnetic 
field.\cite{jarillo,duke1}

In this work, we consider transport through CNT QDs,
focussing on their behavior in a magnetic field.
We analyze the system's evolution from the ultraviolet (UV) to the
infrared (IR) fixed points (FPs); we discuss the KEs that arise and their 
consequences.  
More specifically, we consider the KEs arising from a 
single electron (referred to as {\sl 1/4-filled}) as well as two 
electrons (referred to as {\sl 1/2-filled}) occupying the energy 
levels of the CNT QD closest to the Fermi energy $E_F$ of the leads. 
While previous works detailed the properties of the 1/4-filled 
QD,\cite{aguado} we show that the KEs arising from the 
1/4-filled and 1/2-filled QDs are drastically different; these 
differences have pronounced observable consequences.
%
% Besides CNT QDs, our results are relevant to double QDs
% and, more generally, to QDs with two-fold orbital degeneracy.

%%%%%%%%%%%%%%%%%%%%%%%%%%%%%%%%%%%%%%%%%%%%%%%%%%%%%%%%%%%%%%%%%%%%%%%%%%%
%%%%%%%%%%%%%%%%%%%%%%%%%%%%%%%%%%%%%%%%%%%%%%%%%%%%%%%%%%%%%%%%%%%%%%%%%%%

% \section{Review of Carbon Nanotubes; Carbon Nanotube "Quantum Dot"}

%%%%%%%%%%%%%%%%%%%%%%%%%%%%%%%%%%%%%%%%%%%%%%%%%%%%%%%%%%%%%%%%%%%%%%%%%%%

% \subsection{Carbon Nanotube Quantum Dot}

In what follows, we will be interested in the system's low-energy 
physics; hence, we focus on the energy levels of the CNT QD closest 
to $E_F$ of the leads.
In the absence of magnetic fields, there are two degenerate energy
levels,\cite{jarillo,orbital} which we label as $\alpha$ and $\beta$.
The Hamiltonian we consider is
\begin{eqnarray}
  & & H_{\rm QD} = \frac{E_C}{2}~ \left( \hat{N} - N_0 \right)^2 
  - \frac{h_0}{2} \sum_s \left( \hat{n}_{\alpha s} - \hat{n}_{\beta s} \right)   
%   \nonumber  \\
   \label{anderson}   \\  
  & & \ \ \ \ + \sum_{\kappa,s} \left( \left[ t_1~ \psi^{\dagger}_{1 \kappa s}(0) 
  + t_2~ \psi^{\dagger}_{2 \kappa s}(0) \right] d^{\phantom \dagger}_{\kappa s} 
     + \ h.c. \right)
%   + d^{\dagger}_{i,s} \psi^{\phantom \dagger}_{i,s}(0) \right) 
  \, ,  
%  \label{anderson}  
\nonumber  
\end{eqnarray}
where $\psi^{\dagger}_{i \kappa s}(0)$ creates an electron (at $x$=0) 
with spin-$s$ in band-$\kappa$ from lead-$i$ ($i$=1,2);
$d^{\dagger}_{\kappa s}$ creates an electron with spin-s 
in orbital-$\kappa$ ($\kappa$=$\alpha$,$\beta$) on the QD; 
$\hat{n}_{\kappa s}$=$d^{\dagger}_{\kappa s} 
  d^{\phantom \dagger}_{\kappa s}$
and $\hat{N}$=$\sum_{\kappa,s}$$\hat{n}_{\kappa s}$;
$N_0$ is the optimal number of electrons on the QD, which 
can be controlled by a gate voltage; $E_C$ is the charging energy; 
$t_i$ is the tunneling matrix element between lead-$i$ and the QD; 
$h_0$ is a magnetic field.
In this work, we take the $\{t_i \}$ to conserve the orbital quantum 
number (which is relevant to the experiments in 
Refs.~\onlinecite{duke1} and \onlinecite{duke2});\cite{dukenrg} 
as a result, the system has an SU(4) symmetry when 
$h_0$=0.\cite{SU4review}
$h_0$, which would arise from a magnetic field applied 
parallel to the CNT's axis, splits the $\alpha$ and $\beta$ 
orbitals.
Throughout this work, we employ units where $\hbar$=1.

It should be noted $h_0$ would also give rise to a Zeeman splitting, but 
this splitting is considerably smaller than the orbital splitting,
particularly for larger diameter CNTs.
Indeed, the orbital moment $\mu_{\rm orb}$ of a 5nm diameter CNT was 
found to be $\mu_{\rm orb}$$\simeq$1.5meV/T\cite{orbital} 
i.e. $\mu_{\rm orb}$$\simeq$26$\mu_B$.  [$\mu_B$ is the Bohr magneton.]
As we will be interested in small fields --- 
$h_0$$\sim$${\cal O}(T_K^{\rm SU(4)})$, where $T_K^{\rm SU(4)}$ is 
given by Eq.~\ref{TK} --- the Zeeman splitting will have very small effects.  
Therefore, in what follows, we focus on the orbital splitting.

%%%%%%%%%%%%%%%%%%%%%%%%%%%%%%%%%%%%%%%%%%%%%%%%%%%%%%%%%%%%%%%%%%%%%%%%%%%
%%%%%%%%%%%%%%%%%%%%%%%%%%%%%%%%%%%%%%%%%%%%%%%%%%%%%%%%%%%%%%%%%%%%%%%%%%%

% \section{From the Mixed Valence to the Kondo Regimes}

We begin our discussion of the properties of CNT QDs by 
considering the current $I$=$\langle$$\hat{I}$$\rangle$, where
$\hat{I}$ is the current operator
\begin{equation}
 \hat{I} = -iet_1 \sum_{\kappa,s} \left[ 
   \psi^{\dagger}_{1 \kappa s}(0) d^{\phantom \dagger}_{\kappa s}  
  - d^{\dagger}_{\kappa s} \psi^{\phantom \dagger}_{1 \kappa s}(0) 
  \right] \ 
\label{current}
\end{equation}
($e$ is the electron's charge); in particular, we compute the 
conductance
$G$=$dI/dV$ vs. $N_0$ (in linear response).
We are interested in the behavior of $G$ as $h_0$ is varied, 
as well as how $G$ evolves (with temperature) from the UV to 
the IR FPs.
To understand the IR behavior, $G$ was computed as per 
Ref.~\onlinecite{landauer} using the logarithmic-descretization 
embedded cluster approximation (LDECA)\cite{ldeca} and the Friedel 
sum rule\cite{kondobook};
to treat the UV regime --- 
$T$$\gg$$\Gamma_i$, where $\Gamma_i$=$2\pi \rho_0 t_i^2$ with 
$\rho_0$ being the electrons' density of states in the leads 
--- we employed a master equation approach.\cite{been}

\begin{figure}[t]
\scalebox{.46}{\hspace{-2.75in} \includegraphics{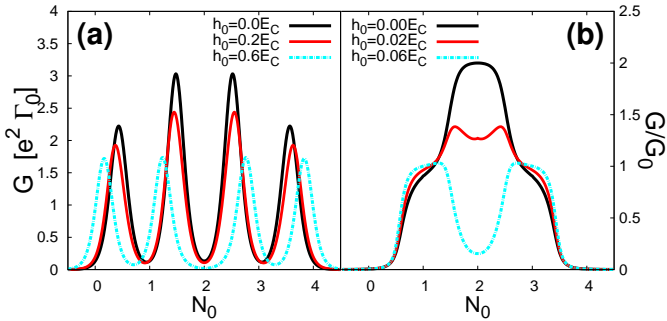} }
\caption{ $G$=$dI/dV$ vs. $N_0$ in linear response for several 
values of $h_0$: (a) $T$=0.1$E_C$ and (b) $T$=0.}
\label{fig:Gvsgate}
\end{figure}

Fig.~\ref{fig:Gvsgate}a shows $G$ vs. $N_0$ in the 
UV regime for several values of $h_0$.
Letting $\Gamma_0$=2$\Gamma_1 \Gamma_2$/($\Gamma_1$+$\Gamma_2$),
\begin{eqnarray}
 G & = & e^2 \Gamma_0  \hspace{-.27in}
  \sum_{\{N_{\alpha},N_{\beta}, N'_{\alpha},N'_{\beta}\} }
  \hspace{-.27in}
  {\rm max}\{ M_{N_{\alpha}N_{\beta}},M_{N'_{\alpha}N'_{\beta}} \}~
  P_{N_{\alpha}N_{\beta}}
 \nonumber \\  & \times &  
  \frac{\exp [(E_{N_{\alpha}N_{\beta}}-E_{N'_{\alpha}N'_{\beta}}
    )/T ] + 1}{ 8T~
  \cosh^2 [(E_{N_{\alpha}N_{\beta}}-E_{N'_{\alpha}N'_{\beta}}
    )/2T ] }  \ ,
 \nonumber
\end{eqnarray}
where $P_{N_{\alpha}N_{\beta}}$ is the probability for the QD to be
in a state with $N_{\alpha}$ ($N_{\beta}$) electrons in the $\alpha$ 
($\beta$) orbital, 
$E_{N_{\alpha}N_{\beta}}$ is the energy of the state with 
$M_{N_{\alpha}N_{\beta}}$ being the number of these states,
and the 
$\{N_{\alpha},N_{\beta}, N'_{\alpha},N'_{\beta}\}$ satisfy
$(N_{\alpha}$+$N_{\beta})$$-$$(N'_{\alpha}$+$N'_{\beta})$=1.
In Fig.~\ref{fig:Gvsgate}a, we observe the well-known CB peaks for 
$N_0$=$N$+1/2 
($N$ is an integer) and valleys for $N_0$=$N$.
When $h_0$=0, the system has an SU(4) symmetry; 
the two middle peaks have more spectral weight 
e.g. the peak at $N_0$=3/2 
(due to fluctuations between states with $N$=1 and $N$=2)  
has more spectral weight than the peak at $N_0$=1/2 
(due to fluctuations between states with $N$=0 and $N$=1).  
From the above expression for $G$, this occurs because there 
are more states with $N$=2 than $N$=1 or $N$=0.
When $h_0$$\neq$0, the SU(4) symmetry is reduced to SU(2); as a 
result, the peaks are split and the spectral weight becomes evenly 
distributed.

Fig.~\ref{fig:Gvsgate}b shows $G$/$G_0$ vs. $N_0$ at $T$=0, where 
$G_0$=($e^2$/$\pi$)4$\Gamma_1$$\Gamma_2$/($\Gamma_1$+$\Gamma_2$)$^2$.
Rather than four peaks, we see three distinct plateaus when $h_0$=0
--- $G$/$G_0$=1 for the plateaus centered about $N_0$=1 and $N_0$=3;
$G$/$G_0$=2 for the plateau centered about $N_0$=2.
Furthermore, $h_0$ has interesting effects on $G$ --- whereas $h_0$
mainly splits the peaks in the UV regime (Fig.~\ref{fig:Gvsgate}a), 
$h_0$ has more drastic effects in the IR.  
Indeed, the plateau centered about $N_0$=2 is suppressed by $h_0$;
the plateaus centered about $N_0$=1 and $N_0$=3, on the other hand, 
are unaffected.
As discussed below, the behavior at $T$=0 occurs because SC states 
between the QD and leads are formed; 
$h_0$ has drastic effects on the SC states.

%%%%%%%%%%%%%%%%%%%%%%%%%%%%%%%%%%%%%%%%%%%%%%%%%%%%%%%%%%%%%%%%%%%%%%%%%%%
%%%%%%%%%%%%%%%%%%%%%%%%%%%%%%%%%%%%%%%%%%%%%%%%%%%%%%%%%%%%%%%%%%%%%%%%%%%

% \section{Numerical Spectral Functions}

\begin{figure}[t]
\scalebox{.47}{\hspace{-2.5in} \includegraphics{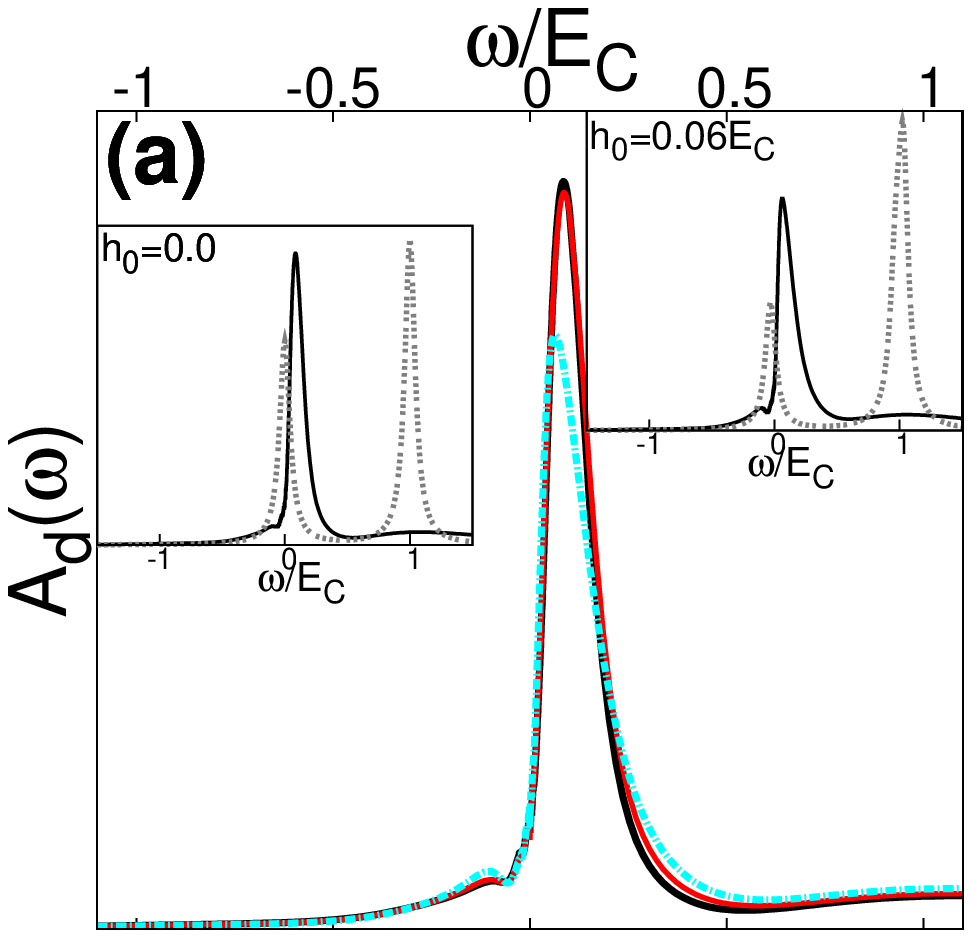} } 

\vspace{-.072in}
\scalebox{.466}{\hspace{-2.8in} \includegraphics{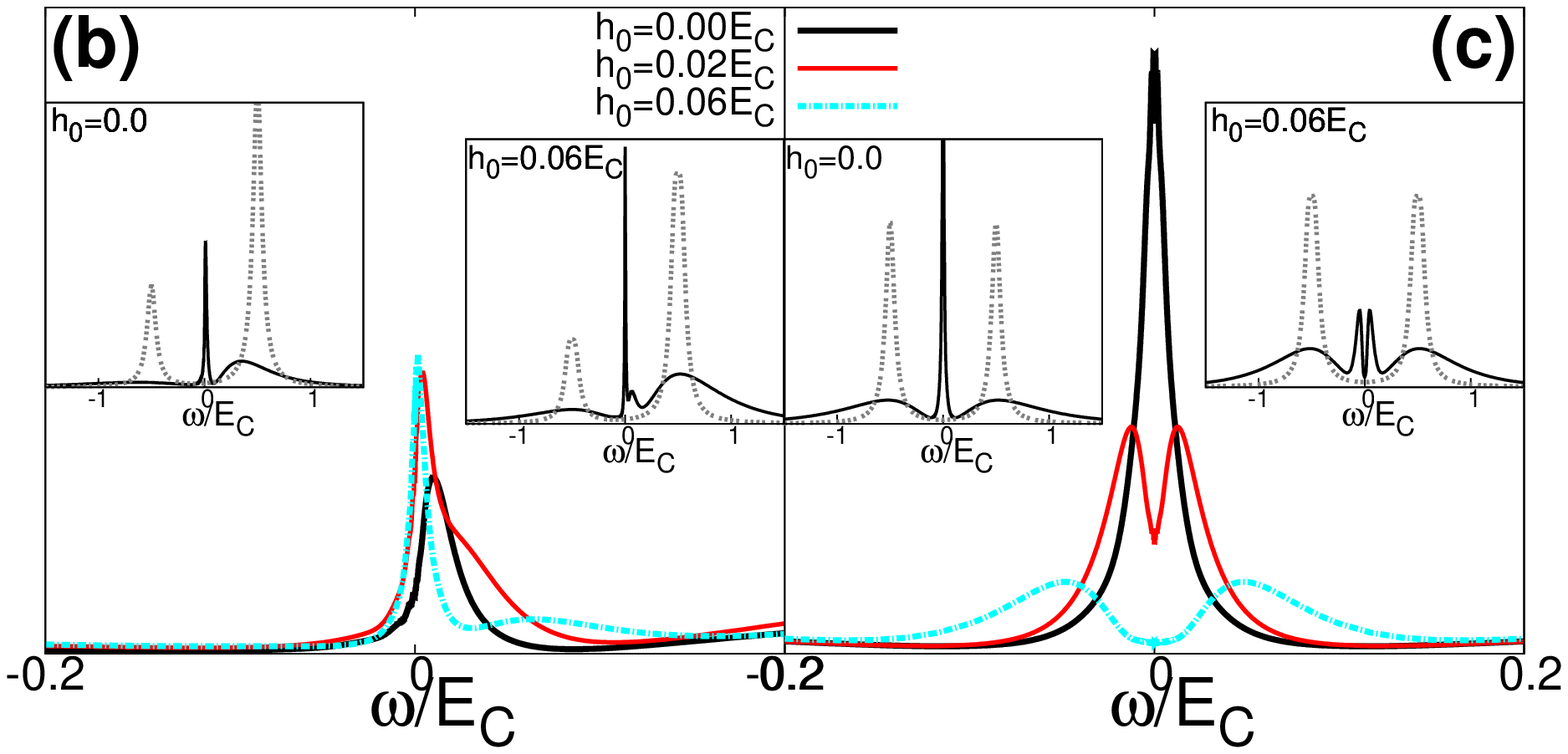} }
\caption{ QD's spectral function $A_d(\omega)$.  
(a) $A_d(\omega)$ for $N_0$=1/2.  (b) $A_d(\omega)$ for $N_0$=1.  
(c) $A_d(\omega)$ for $N_0$=2.
Insets: Comparison of $A_d(\omega)$ at the UV (gray dotted lines) 
and IR (solid black lines) fixed points.}
\label{fig:spectral}
\end{figure}

We now address the physics behind Fig.~\ref{fig:Gvsgate} ---
we investigate the SC states which arise in the IR, as well 
as how they evolved from the UV FP.
To this end, we examine the QD's spectral function (SF), 
$A_d(\omega)$.
Fig.~\ref{fig:spectral} shows $A_d(\omega)$
(at $T$=0) obtained via the LDECA.  For comparison, results for 
$A_d(\omega)$ at the UV FP --- obtained by formally setting 
\{$t_i$\}=0 --- are shown in the insets.
Fig.~\ref{fig:spectral}a shows $A_d(\omega)$ at the $N_0$=1/2 CB peak.  
Here we see a broad peak near $\omega$=0 i.e. near $E_F$; 
its features do not change much with $h_0$.  
From the insets, we see there was a redistribution of spectral weight, 
with much of the peak's weight in the IR having been transferred from 
higher energies.

Figs.~\ref{fig:spectral}b and \ref{fig:spectral}c show $A_d(\omega)$ 
in the CB valleys.
A key feature is the narrow resonance which appears
at or near $E_F$ --- the Kondo resonance (KR).  This resonance is a 
consequence of the SC state formed between the QD and leads due to
the KE; its width represents the dynamically generated scale 
characteristic of the SC state --- the Kondo temperature, $T_K$.
As discussed below, the position and width of the KR are 
characteristic of the particular Kondo fixed point (KFP).

Fig.~\ref{fig:spectral}b shows $A_d(\omega)$ in the $N_0$=1 
valley i.e. the 1/4-filled QD.  
For $h_0$=0, $A_d(\omega)$ exhibits a KR near $E_F$; 
for $h_0$$\neq$0, the resonance moves toward $E_F$ and its width
narrows.
As mentioned above, when $h_0$=0 the system has an SU(4) symmetry;
$h_0$$\neq$0 reduces this symmetry to SU(2).
For the 1/4-filled QD, the system flows to the SU(4) KFP when 
$h_0$=0, while $h_0$$\neq$0 drives the system to the SU(2) KFP; 
the KR is near (at) $E_F$ at the SU(4) (SU(2)) KFP with 
$T_K^{\rm SU(2)}$$<$$T_K^{\rm SU(4)}$.
The UV and IR behaviors of $A_d(\omega)$ are compared in the
insets --- the KR is indeed an IR property, with its
spectral weight taken from the higher energy UV peaks;
interestingly, $h_0$ does not change the qualitative features 
of $A_d(\omega)$ at either the UV or IR FPs.

Fig.~\ref{fig:spectral}c shows $A_d(\omega)$ for $N_0$=2 i.e. the 
1/2-filled QD --- its behavior is drastically different from 
the SFs arising for both $N_0$=1/2 and $N_0$=1.
For $h_0$=0, $A_d(\omega)$ exhibits a narrow KR at 
$E_F$; for $h_0$$\neq$0, the resonance splits and is suppressed.
Hence, contrary to the 1/4-filled QD where $h_0$ drives the system
from the SU(4) KFP to the SU(2) KFP, $h_0$ destroys the KE for the 
1/2-filled QD.
The UV and IR behaviors of $A_d(\omega)$ are compared in the insets 
--- we see the KR suppressed as $h_0$ increases; as this
occurs, the peaks at $\omega$=$\pm$$E_C$/2 regain spectral weight.

%%%%%%%%%%%%%%%%%%%%%%%%%%%%%%%%%%%%%%%%%%%%%%%%%%%%%%%%%%%%%%%%%%%%%%%%%%%
%%%%%%%%%%%%%%%%%%%%%%%%%%%%%%%%%%%%%%%%%%%%%%%%%%%%%%%%%%%%%%%%%%%%%%%%%%%

% \section{Effective Kondo Model; "Drone Fermion" Analysis}

Having discussed the QD's SF in the various 
regimes, we now discuss (further) consequences of the 
SF's features in the CB valleys 
i.e. for $N_0$$\simeq$$N$.
To facilitate the analysis, we integrate out charge fluctuations 
on the QD; we arrive at the Coqblin-Schrieffer Hamiltonian\cite{kondobook}
\begin{equation}
  H_{\rm QD} = - \frac{J}{4} \left( \psi^{\dagger}_{\kappa s\phantom'}
     f^{\phantom \dagger}_{\kappa s} \right) \left( 
     f^{\dagger}_{\kappa' s'} \psi^{\phantom \dagger}_{\kappa' s'} 
  \right) - \frac{h_0}{2}~ f^{\dagger}_{\kappa s}  
    \sigma^{z}_{\kappa \kappa'} f^{\phantom \dagger}_{\kappa' s} 
\label{schreifferwolff} 
\end{equation}
where $\psi^{\phantom \dagger}_{\kappa s}$=$[t_1 \psi^{\phantom 
\dagger}_{1\kappa s}(0)+t_2\psi^{\phantom \dagger}_{2\kappa s}(0)]
$/$t$ with $t$=$\sqrt{t_1^2 + t_2^2}$,
$J=(4t^2/E_C)\left[(N-N_0-1/2)^{-1}-(N-N_0+1/2)^{-1}\right]$,
and the fermion operators satisfy the constraint 
$f^{\dagger}_{\kappa s} f^{\phantom \dagger}_{\kappa s}$=$N$
with $N$ being the number of particles on the QD.
(While discussing the physics of the CB valleys, we write the
QD's fermion operators as $\{ f_{\kappa s} \}$; also, Einstein 
summation convention is utilized).
To treat Eq.~\ref{schreifferwolff}, we consider a path 
integral representation of the partition function -- 
we enforce the constraint
 $f^{\dagger}_{\kappa s} f^{\phantom \dagger}_{\kappa s}$=$N$ 
with a Lagrange multiplier field $\lambda$;
we decouple the Kondo interaction using a Hubbard-Stratonovich 
field $\chi$.\cite{kondobook}  We arrive at an effective 
Hamiltonian
\begin{eqnarray}
 & & H_{\rm eff} = - \frac{h_0}{2}~ f^{\dagger}_{\kappa s}  
     \sigma^{z}_{\kappa \kappa'} f^{\phantom \dagger}_{\kappa' s}
  + \lambda \left( f^{\dagger}_{\kappa s} 
      f^{\phantom \dagger}_{\kappa s} - N \right) 
 \label{Heff}  \\
  & & \ \ \ \ \ \  + \ \frac{4}{J} |\chi|^2 
   + \chi^{\dagger} f^{\dagger}_{\kappa s} 
        \psi^{\phantom \dagger}_{\kappa s}
  + \chi~ \psi^{\dagger}_{\kappa s} f^{\phantom \dagger}_{\kappa s}  
  \  .  \nonumber
% \label{Heff}
\end{eqnarray}

%%%%%%%%%%%%%%%%%%%%%%%%%%%%%%%%%%%%%%%%%%%%%%%%%%%%%%%%%%%%%%%%%%%%%%%%%%%

% \subsubsection{Weak Coupling Fixed Point}

\begin{figure}[b]
\scalebox{.30}{\includegraphics{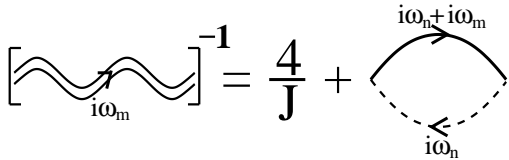} }
\caption{$\chi$ field propagator ${\cal J}(i\omega_m)$ 
--- the solid (dashed) line denotes the leads' ($f$-fermions') 
Green's function.}
\label{fig:diagram}
\end{figure} 

We begin by considering the physics at higher energies, 
focussing on the flow from the UV to the IR FPs.
To do so, we treat the Bose fields $\chi$ and $\lambda$ in
Eq.~\ref{Heff} in mean-field theory (MFT).
Treating $\lambda$ in MFT amounts to treating 
the constraint 
$f^{\dagger}_{\kappa s} f^{\phantom \dagger}_{\kappa s}$=$N$ 
on average: 
$\langle f^{\dagger}_{\kappa s} 
 f^{\phantom \dagger}_{\kappa s} \rangle$=$N$.
To describe the physics near the UV FP, we 
take $\langle \chi \rangle$=0;
the physics of the KE is contained in the effective action for 
$\chi$, obtained by integrating out the $f$-fermions and leads.
To one-loop order, the propagator of the $\chi$ field 
${\cal J}(i\omega_m)$ is given by the diagram in 
Fig.~\ref{fig:diagram} (with $\omega_m$ being a boson Matsubara 
frequency).
Physically, ${\cal J}(i\omega_m)$ is the running Kondo 
coupling.\cite{coleman}

\begin{figure}[t]
\scalebox{.5}{\hspace{-2.09in} \includegraphics{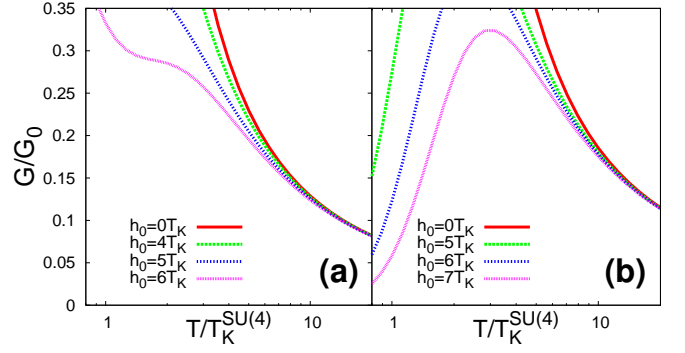} }
\caption{ $G$/$G_0$ vs. $T$/$T_K^{\rm SU(4)}$ in linear response 
in the CB valleys: (a) 1/4-filled QD and (b) 1/2-filled QD.}
\label{fig:GvsT}
\end{figure}

Using our result for ${\cal J}(i\omega_m)$, the current 
$I$=$\langle \hat{I} \rangle$ 
was computed as per Ref.~\onlinecite{landauer};
results for $G$/$G_0$ vs. $T$/$T_K^{\rm SU(4)}$ 
are shown in Fig.~\ref{fig:GvsT}, where 
\begin{equation}
 T_K^{\rm SU(4)} = D \exp \left( -1/\rho_0 J \right) 
\label{TK}
\end{equation}
with $D$ being half the leads' bandwidth.
[As before, 
$G_0$=($e^2$/$\pi$)4$\Gamma_1$$\Gamma_2$/($\Gamma_1$+$\Gamma_2$)$^2$.]
Fig.~\ref{fig:GvsT}a shows results for the 1/4-filled QD.
To begin with, we see that $G$ grows logarithmically as $T$ 
is reduced --- this is a consequence of the logarithmic growth 
of the running Kondo coupling.\cite{kondobook}
Furthermore, $G$/$G_0$ always grows to ${\cal O}(1)$ 
i.e. the system always flows to strong coupling.  This is because 
the 1/4-filled QD exhibits a KE, irrespective of the 
value of $h_0$.
However, $G$ grows more slowly for larger $h_0$ --- the system flows 
to the SU(4) (SU(2)) KFP for smaller (larger) $h_0$; the slower growth 
of $G$ for larger $h_0$ occurs because $T_K^{\rm SU(2)}$$<$$T_K^{\rm SU(4)}$.
(See Fig.~\ref{fig:spectral}b.)

Fig.~\ref{fig:GvsT}b shows $G$/$G_0$ vs. $T$/$T_K^{\rm SU(4)}$ 
for the 1/2-filled QD; 
the results are drastically different from the 1/4-filled QD.  
As discussed above, whereas $h_0$ drives the system from the 
SU(4) to the SU(2) KFP for the 1/4-filled QD, $h_0$ destroys 
the KE for the 1/2-filled QD (see Fig.~\ref{fig:spectral}c);
$G$ even becomes non-monotonic.   
Such behavior has been observed in magnetic alloys, where the 
occurrence of a spin glass phase freezes spin-flip processes 
and, hence, suppresses the KE.\cite{spinglass}  
Here, larger values of $h_0$ freeze both spin and orbital processes.
More precisely, $h_0$ cuts off the growth of the running Kondo coupling 
--- for the 1/2-filled QD ${\cal J}$($T$)$\equiv$${\cal J}$($i\omega_m$=0) 
is given by 
\begin{equation}
 \rho_0 {\cal J}(T) = J \left\{ 
   \ln \left(\frac{2\pi T}{T_K^{\rm SU(4)}} \right) 
 + {\rm Re}\left[\psi \left(\frac{1}{2} + i \frac{h_0}{4\pi T} \right) 
   \right] \right\}^{-1}  \hspace{-.123in}  ,
\nonumber 
\end{equation}
where $\psi(z)$ is the digamma function;\cite{gradshteyn}
as $\psi(z)$$\simeq$$\ln(z)$ for $|z|$$\gg$1, 
the growth of ${\cal J}$($T$) is suppressed for $h_0$ sufficiently larger 
than $T_K^{\rm SU(4)}$.

%%%%%%%%%%%%%%%%%%%%%%%%%%%%%%%%%%%%%%%%%%%%%%%%%%%%%%%%%%%%%%%%%%%%%%%%%%%

% \subsubsection{Strong Coupling Fixed Point}

Having discussed the flow (in the CB valleys) from the UV to the 
KFPs in the IR, we now discuss further the physics of the SC KFPs.
As before, we treat the Bose fields in Eq.~\ref{Heff} in MFT.  
Now, however, to describe the physics of the SC KFPs, we take 
$\langle \chi \rangle$=$\chi_0$($\neq$0).\cite{kondobook}  
Hence, $\lambda$ and $\chi_0$ are determined via
$\langle f^{\dagger}_{\kappa s} 
  f^{\phantom \dagger}_{\kappa s} \rangle$=$N$ and
$\chi_0$+$2 J \langle \psi^{\dagger}_{\kappa s} 
 f^{\phantom \dagger}_{\kappa s} \rangle$=0.
With $\langle \chi \rangle$$\neq$0, the $f$-fermions SF is 
\begin{equation}
 A^f_i(\omega) = \frac{2\Gamma}
   {(\omega - \varepsilon_i)^2 + \Gamma^2}  \  
\nonumber
\end{equation}
($\varepsilon_{1/2}$=$\lambda$$\pm$$h_0/2$), 
where $\Gamma$=$T_K$ when $T$=0.\cite{kondobook}

\begin{figure}[t]
\scalebox{.69}{\includegraphics{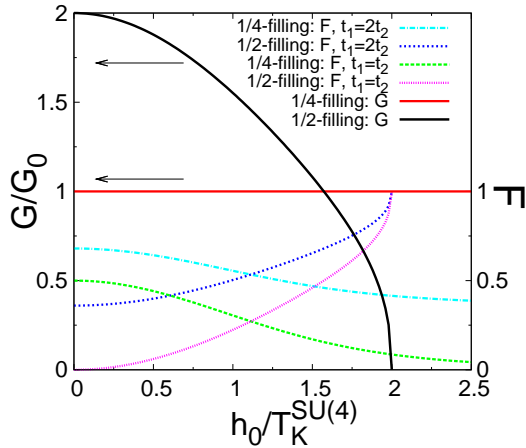} }
\caption{ $G$/$G_0$ vs. $h_0$/$T_K^{\rm SU(4)}$ and 
$F$ vs. $h_0$/$T_K^{\rm SU(4)}$ in linear response at $T$=0. }
\label{fig:Gvsh}
\end{figure}

Fig.~\ref{fig:Gvsh} shows $G$/$G_0$ vs. $h_0$/$T_K^{\rm SU(4)}$ 
(computed as per Ref.~\onlinecite{landauer}) at $T$=0.
For the 1/4-filled QD, $G$/$G_0$=1 regardless of the value of $h_0$.
This occurs because there is always a KE, $\Gamma$$\neq$0 ---
for small $h_0$, one is in the SU(4) Kondo regime; for larger $h_0$,
one crosses over to the SU(2) Kondo regime.
For the 1/2-filled QD, on the other hand, we see that $G$ depends 
on the magnitude of $h_0$ --- $G$/$G_0$=2 for $h_0$=0 and decreases 
as $h_0$ is increased.
[Within MFT, $G$$\rightarrow$0 for $h_0$=2$T_K^{\rm SU(4)}$.]
This is because the SU(4) KE is destroyed and, consequently, the KR
in the QD's SF is suppressed for $h_0$ sufficiently large.
(See Fig.~\ref{fig:spectral}c.)

%%%%%%%%%%%%%%%%%%%%%%%%%%%%%%%%%%%%%%%%%%%%%%%%%%%%%%%%%%%%%%%%%%%%%%%%%%% 
   
% \section{Noise in the Kondo Regime}

Also shown in Fig.~\ref{fig:Gvsh} are results for the noise
(which has been shown to be a powerful probe of Kondo 
physics\cite{kondonoise,kondonoiseexp})
at $T$=0.
More specifically, we computed the zero-frequency noise
\begin{equation}
 S(eV) = \int dt~ \left[ \langle \hat{I}(t) \hat{I} \rangle 
       - \langle \hat{I} \rangle^2 \right]  
\end{equation}
% ($\hat{I}$ is from Eq.~\ref{current})
and, subsequently, the Fano factor $F$$\equiv$$S/2eI$ in
linear response.
While the conductance probes the spectral weight of the KR, 
the noise gives information about its position.
Indeed, while both the SU(4) and SU(2) KEs give $G$/$G_0$=1
for the 1/4-filled QD, differences between the two can 
drastically be seen in $F$ ---
$F$ decreases as $h_0$ increases i.e. as we move from the SU(4)
to the SU(2) KFP.
This is seen most dramatically when $t_1$=$t_2$ where 
$F$$\rightarrow$0 as $h_0$ increases, but the qualitative
behavior of $F$ is robust.  [See the results for 
$t_1$=2$t_2$.]
Physically, this arises because the KR in the QD's 
SF is at $E_F$ at the SU(2) KFP, while it is 
away from $E_F$ at the SU(4) KFP.  (See Fig.~\ref{fig:spectral}b.)
For the 1/2-filled QD, $F$ increases as $h_0$ increases, 
approaching unity as $G$$\rightarrow$0;
for $t_1$=$t_2$, $F$$\rightarrow$0 as $h_0$$\rightarrow$0.
For $h_0$=0, the system is in a SC state, with a KR at $E_F$; as 
a result $F$=0 when $t_1$=$t_2$.  
As $h_0$ is increased, the SC state is destroyed and the system is 
driven to the weak-coupling regime; hence, $F$$\rightarrow$1 
i.e. $F$ becomes Poissonian.\cite{noisereview}

%%%%%%%%%%%%%%%%%%%%%%%%%%%%%%%%%%%%%%%%%%%%%%%%%%%%%%%%%%%%%%%%%%%%%%%%%%%
%%%%%%%%%%%%%%%%%%%%%%%%%%%%%%%%%%%%%%%%%%%%%%%%%%%%%%%%%%%%%%%%%%%%%%%%%%%

% \section{Concluding Remarks}
  
To summarize, we considered the behavior of CNT QDs
in a magnetic field.  We analyzed the evolution of the 
system from the UV to the IR FPs.  We discussed the KEs 
that occur and their experimental consequences.
In particular, the KEs arising for the 1/4-filled and
1/2-filled QDs were shown to be drastically different,
with distinct signatures in the system's transport;
we are optimistic our results, particularly for the 
noise,\cite{noisecomment} can be observed experimentally.
Besides CNT QDs, our results are relevant to double QDs
and, more generally, to QDs with two-fold orbital degeneracy.

%%%%%%%%%%%%%%%%%%%%%%%%%%%%%%%%%%%%%%%%%%%%%%%%%%%%%%%%%%%%%%%%%%%%%%%%%%%
%%%%%%%%%%%%%%%%%%%%%%%%%%%%%%%%%%%%%%%%%%%%%%%%%%%%%%%%%%%%%%%%%%%%%%%%%%%

% \section*{Acknowledgements}

This work was supported by the NSERC of Canada (MM and EHK), 
a SHARCNET Research Chair (MM and EHK), 
and the NSF (DMR - 0710529) (GBM). 

%%%%%%%%%%%%%%%%%%%%%%%%%%%%%%%%%%%%%%%%%%%%%%%%%%%%%%%%%%%%%%%%%%%%%%%%%%%
%%%%%%%%%%%%%%%%%%%%%%%%%%%%%%%%%%%%%%%%%%%%%%%%%%%%%%%%%%%%%%%%%%%%%%%%%%%

%%%%%%%%%%%%%%%%%%%%%%%%%%%%%%%%%%%%%%%%%%%%%%%%%%%%%%%%%%%%%%%%%%%%%%%%%%%
%%%%%%%%%%%%%%%%%%%%%%%%%%%%%%%%%%%%%%%%%%%%%%%%%%%%%%%%%%%%%%%%%%%%%%%%%%%


\begin{thebibliography}{100}
\bibitem{noisecomment}{Indeed, by working at lower temperatures 
and in a magnetic field, we are optimistic our results for the
noise could be observed in the device/setup of 
Ref.~\onlinecite{kondonoiseexp}.}
\end{thebibliography}
\end{document}